# A Data-Driven Machine Learning Approach for Consumer Modeling with Load Disaggregation

A. Khaled Zarabie, Sanjoy Das*, and Hongyu Wu

Electrical & Computer Engineering, Kansas State University.
*Corresponding author: sdas@ksu.edu

*Abstract*—While non-parametric models, such as neural networks, are sufficient in the load forecasting, separate estimates of fixed and shiftable loads are beneficial to a wide range of applications such as distribution system operational planning, load scheduling, energy trading, and utility demand response programs. A semi-parametric estimation model is usually required, where cost sensitivities of demands must be known. Existing research work consistently uses somewhat arbitrary parameters that seem to work best. In this paper, we propose a generic class of data-driven semiparametric models derived from consumption data of residential consumers. A two-stage machine learning approach is developed. In the first stage, disaggregation of the load into fixed and shiftable components is accomplished by means of a hybrid algorithm consisting of non-negative matrix factorization (NMF) and Gaussian mixture models (GMM), with the latter trained by an expectation-maximization (EM) algorithm. The fixed and shiftable loads are subject to analytic treatment with economic considerations. In the second stage, the model parameters are estimated using an $L_2$-norm, $\epsilon$-insensitive regression approach. Actual energy usage data of two residential customers show the validity of the proposed method.

## I. INTRODUCTION

THE overarching goal of this paper is to develop a framework to model a residential unit's daily energy consumption patterns, factoring in cost and temperature dependencies, using non-intrusive smart meter data. Load disaggregation lies at the core of the proposed approach.

Unlike earlier methods that attempt to extract usages of individual appliances, disaggregation in this research separates the load into only two components, (*i*) *shiftable loads*, and (*ii*) *fixed loads*, with only the latter responsive to time-of-use (TOU) pricing and ambient temperatures. The proposed unsupervised learning method for load disaggregation is a *hybrid* algorithm that applies expectation-maximization (EM) to train a Gaussian mixture model (GMM), with non-negative matrix factorization (NMF).

A meaningful, nonlinear utility model of the residence, which is grounded on classical econometrics and game theory, is proposed here. The disaggregated loads are used to estimate the parameters of this model. Regression with an $\epsilon$-insensitive, $L_2$-norm loss function is applied to minimize the mismatch between the model's predicted energy use and the corresponding disaggregated loads.

### A. Literature Survey

Recent advancements in machine learning have made it possible to disaggregate residential loads obtained from smart meters into smaller components. Some algorithms for load disaggregation rely on supervised or semi-supervised learning, which require a set of training data where the individual loads are known a priori [1]. Deep neural networks are an example of such a method [2], [3], [4], [5]. Unfortunately, these approaches are not suitable for non-intrusive load monitoring where the only available data is in the form of smart meter readings, even during the training phase.

Several other approaches tailored for energy disaggregation have been proposed. Among them include [6] that uses the fuzzy c-means clustering to identify the number of appliances. Quadratic programming has been used in [7]. A computationally efficient additive neural network is used in [8], with a specialized training algorithm called cogent confabulation. The approach in [8] relies on OFF-ON transitions to detect individual appliances. It also uses bagging (an ensemble learning technique) for improved classification of appliances. Two novel algorithms based on spectral graph theory have been applied for disaggregation in [9] whereas fuzzy logic has been adopted in [10].

Hidden Markov models (HMM) are a popular choice of modeling individual appliances. Individual appliances are represented using two or more states, with transition and/or emission probabilities usually learned from a variation of the EM algorithm. In [11], the total number of appliances are determined using *K*-means clustering. An improved method proposed in [12] that models appliances using factorial HMMs. Other extensions to represent appliances include explicit HMMs [13] and hierarchical HMMs [14] for load modeling.

The EM algorithm is also used to train Gaussian mixture models (GMM) [15]. GMMs are based on probability theory that usually apply the maximum likelihood criterion for appliance classification. These approaches assume the presence of a set of 'latent' variables $z_j$, where $j$ is an index, with only one being 'active' at each time. The output follows a Gaussian probability distribution that is uniquely identified by the active latent variable. The algorithm proposed in [15] is based on GMMs. The approach uses the Dempster-Shaffer theory for appliance classification.

Non-negative matrix factorization (NMF) is another widely used, unsupervised learning approach for energy disaggregation. The classical NMF algorithm decomposes an input data matrix **X**, whose columns are sample vectors, into two factors, **W** and **H**, so that their product equals **X**. Usually,

**X** has a very large number of columns, which are independent samples. The columns of **W** serve as basis vectors so that each sample $\mathbf{x}(n), n \in \mathcal{N}$, which are columns of **X** can be represented as a weighted combination of the bases, with the non-negative weights being the corresponding column vectors $\mathbf{h}(n)$ of **H**. NFM has been used for energy disaggregation of HVAC load components in an industrial building and in a smart home setting [16]. Another method has been proposed in [17], [18] to impose $L_0$ constraints, which uses a softmax distribution for the elements in **H** to assign weights to them in such a manner that those with higher values are likelier to improve the objective function. Semi-supervised NMF using prior knowledge of the usage time profiles of individual appliances has been proposed in [1]. NMF has been applied for data over larger periods in time to retrieve seasonal trends in usage profiles in [19].

Disaggregated loads are used in this research to derive nonlinear utility models of residential consumers. In [20], generation side utilities are modeled as piecewise linear functions so that linear programming can be applied for energy auctioning. A linear utility function is also used in [21] for individual appliances, with customers trying to maximize the sum of all such utilities. However, it should be noted that linearization is not suitable in many other applications as they do not encapsulate the dependence of utility to price changes. For this reason, recent research proposals routinely represent utilities using saturating nonlinearities. In [22] customers' behavior based utility functions are modeled in this manner. In the double auction mechanism described in [23] participants' bidding strategies are determined from nonlinear utilities of individual goods. Similarly, in [24], nonlinear functions have been used to quantify the utility of a potential user from pricing and QoS.

Logarithmic functions are commonly used to model utilities nowadays. The double auctions in [23] make use of such utilities. Portfolio optimization using power-log utilities are taken up in [25]. The approach in [24] uses log-concave utility functions. In [26], logarithmic utilities are also used to quantify the benefit derived from operating shiftable appliances. In [27], [28], [29] residential consumers' utilities have been modeled as logarithms in energy market games. In [28], log utilities have been used to represent aggregator payoffs. Logarithmic utilities have also beeen used in [30], [31], in a Stackelberg game based model of energy trade in the grid.

### B. Research Contributions

The novelties of the research that this paper entails are listed as follows.

(*i*) A fully non-intrusive approach for load disaggregation is proposed. This is an improvement not only over supervised learning approaches but also those that use unsupervised learning such as GMMs, HMMs, or NMF, most of which require some form of prior information about the consumption patterns of individual appliances. Therefore, the proposed approach can be entirely trained in real-time after deployment.

(*ii*) Disaggregation in this research is accomplished with a hybrid algorithm that harnesses the synergy of GMMs and NMF. GMMs are most effective in learning the load patterns of binary OFF-ON shiftable appliances as well as those operating at only a few discrete levels, whereas NMF approaches are suitable to iteratively obtain basis sets to represent fixed loads whose energy consumption levels cannot be discretized readily.

(*iii*) A semi-parametric consumer utility model has been proposed, which represents the appropriate tradeoff between parametric and non-parametric models. For many applications, it suffices to distinguish only between fixed and cost dependent components of the load. Additionally, the present utility model considers temperature dependence of energy consumption.

(i*v*) Parameter estimation of the consumer utility model has been formulated in terms of $L_2$-norm maximum margin regression. The use of the utility model as a means to validate the performance of the disaggregation algorithm has been proposed.

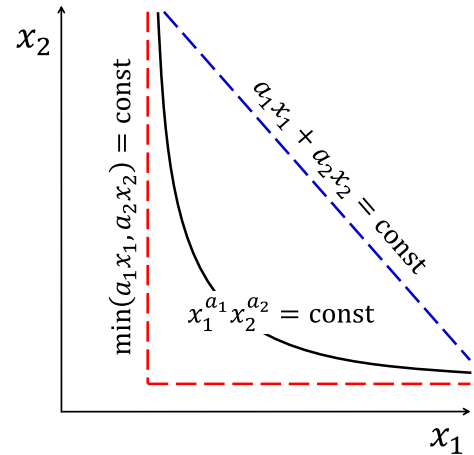

**Fig. 1.** Isoquants for Cobb-Douglas preference (black), perfect substitution (blue, dashed) and perfect complementarity (red, dashed). The coordinates represent resource usages.

## II. OVERVIEW OF APPROACH

It is assumed that any user's energy usage $x_t$ at any time instant $t$ is divided into two components, i.e. $x_t = x_t^f + x_t^s + \eta_t$, where $x_t^f$ and $x_t^s$ are the fixed and shiftable loads, and $\eta_t$ is random noise. We define the following vectors, $\mathbf{x} = [x_t]_{t \in \mathcal{T}}$, $\mathbf{x}^f = [x_t^f]_{t \in \mathcal{T}}$, and $\mathbf{x}^s = [x_t^s]_{t \in \mathcal{T}}$. Each quantity is a $|\mathcal{T}| \times 1$ vector, where $\mathcal{T}$ is the set of all time instances. The time instances $t \in \mathcal{T}$ are divided into a set of non-overlapping periods, with each period including exactly $L$ instances. The set of periods is denoted as $\mathcal{K}$ so that $L = |\mathcal{K}|^{-1}|\mathcal{T}|$. For example, if the time instances are of a minute duration each, that are divided up into hour-long periods, then $|\mathcal{T}| = 1440$ mins/day, $|\mathcal{K}| = 24$ hours/day, and $L = 60$ mins/hour. Within each period $k \in \mathcal{K}$ we define the $L \times 1$ vectors $\mathbf{x}_k^f = [x_t^f]_{\hbar(t)=k}$, $\mathbf{x}_k^s = [x_t^s]_{\hbar(t)=k}$, where $\hbar(\cdot)$ is a mapping $\hbar : \mathcal{T} \to \mathcal{K}$ that provides the period index corresponding to a time instance $t$. The temperature vector $\boldsymbol{\theta} = [\theta_k]_{k \in \mathcal{K}}$ and the vector of unit costs $\mathbf{c} = [c_k]_{k \in \mathcal{K}}$ are the other $L \times 1$ vectors used here.

The dataset used in this research consists of multiple samples, where each sample pertains to a day. Accordingly, we define the load matrices, $\mathbf{X} = [\mathbf{x}(n)]_{n\in\mathcal{N}}$, $\mathbf{X}^f = [\mathbf{x}^f(n)]_{n\in\mathcal{N}}$, and $\mathbf{X}^s = [\mathbf{x}^s(n)]_{n\in\mathcal{N}}$, of dimensionalities $|\mathcal{T}| \times |\mathcal{N}|$, where $\mathcal{N}$ is the set of samples and sample index $n$ indicates a day. The $|\mathcal{K}| \times |\mathcal{N}|$ matrix of temperatures is $\mathbf{\Theta} = [\mathbf{\theta}(n)]_{n\in\mathcal{N}}$. The cost vector $\mathbf{c}$ is constant across all samples in $\mathcal{N}$.

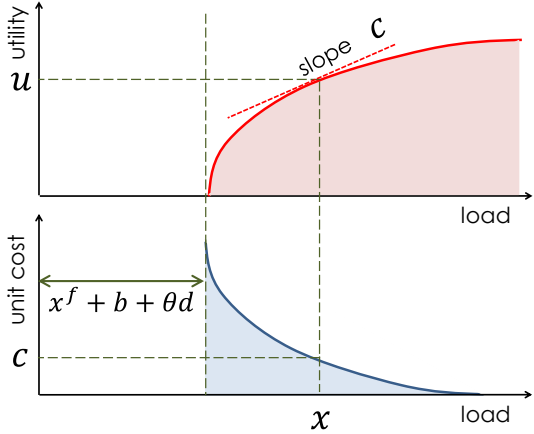

**Fig. 2.** A typical utility function (top) and its derivative, the marginal utility (bottom). The marginal utility is also the unit cost of energy.

*A. Load Model*

Appliances that are used intermittently by a typical residence make up the shiftable loads, such as washer/dryer units or PHEVs. When in operation they consume a large amount of energy that can be discretized into a few levels. Smart homes schedule these appliances based on the instantaneous cost of electricity. Some shiftable loads (e.g. air-conditioners) are temperature dependent, their use being comfort-driven. Fixed loads pertain to appliances that are deemed essential in any residence, such as lighting and refrigeration. Consequently energy pricing has little bearing on their operation. With the exception of refrigerators, fixed loads cannot be characterized as discrete loads. Their energy consumption is lower than that in fixed loads. Unlike shiftable loads, fixed load appliances are operated throughout the day.

It is assumed that the fixed load at any period $k$ consists of a temperature independent and dependent terms so that,
$$\hat{x}_k^f = p_k + q_k \theta_k. \quad (1)$$
Here $p_k$ and $q_k$ are two model parameters, $x_k^f = \mathbf{1}_L^T \mathbf{x}_k^f$ is the total fixed load and $\theta_k$ is the temperature during that period. It may be noted that $q_k = 0$ in most cases as fixed loads do not depend on outside temperature. The 'hat' (ˆ) appearing over $x_k^f$ in (1) indicates that it is a *modeled* load (as opposed to a real load obtained from the available input data). Shiftable loads are modeled using econometric principles that are described next.

Consider a market with two divisible resources, indexed 1 and 2. Suppose the amounts of each resource consumed by a consumer are $x_1$ and $x_2$. The consumer's preference can be quantified in terms of the function $y(x_1, x_2)$. When the resources display perfect complementarity, the consumer's preference can be expressed as $y(x_1, x_2) = \min(a_1 x_1, a_2 x_2)$, where $a_1$ and $a_2$ are two constants based on the consumer's individual characteristics. On the other hand, when the resources are perfect substitutes of each other, then $y(x_1, x_2) = a_1 x_1 + a_2 x_2$. The Cobb-Douglas function [32] is frequently used as the preference when the two resources are intermediate between being perfect complements and perfect substitutes. The schematic in Fig. 1. shows the shape of a typical isoquant for the Cobb-Douglas preference. For comparison, isoquants for perfect substitutes and perfect complements are also provided. In the present case with two resources, the preference can be expressed as $y(x_1, x_2) = x_1^{a_1} x_2^{a_2}$. In general, with $\mathbf{x}$ representing the vector of all resources, the Cobb-Douglas preference is $y(\mathbf{x}) = \prod_k (x_k)^{a_k}$, $(k = 1,2,...)$. The parameters $a_k > 0$ are referred to as elasticity constants. The Cobb-Douglas preference is widely used to model resource usage.

Since the original Cobb-Douglas preference function serves as a production function, it becomes zero if any component of the final product, $x_k = 0$. In the present application, each factor is incremented by 1 to prevent this from happening, so that each factor in $y(\mathbf{x}^s)$ is of the form, $(x_k^s + 1)^{a_k}$. The logarithm of the modified Cobb–Douglas preference is chosen as the utility function in this research. Treating the logarithm of the Cobb-Douglas preference in this manner, renders the utility function: (*i*) non-decreasing, (*ii*) strictly concave, (*iii*) intersecting at the origin at $\mathbf{x}^s = \mathbf{0}$, indicating that a consumer can glean no utility without consuming energy, and (*iv*) Lipschitz continuous. These are highly desirable features of the utility function, the first three being mandated by the well-known economic precept known as the *law of diminishing returns*, while the fourth making the function suitable for mathematical treatment.

Accordingly, the expression for the utility used in this research is given by the following,
$$u(\mathbf{x}^s) = \sum_{k \in \mathcal{K}} a_k \log\left(\left(x_k^s - (b_k + d_k \theta_k)\right)_+ + 1\right). \quad (2)$$
The quantity $b_k$ is the base component of the shiftable load that is determined by extraneous, non-economic factors (e.g. it is not practicable for a washer/dryer unit to be put into use at 2:00 AM). The term $d_k \theta_k$ is the temperature dependent component of $x_k^s$ (which is predominantly due to air conditioning). For this reason, $(b_k + d_k \theta_k)$ is subtracted from $x_k^s$ in (2). The operator $(\cdot)_+$ restricts the enclosed argument to be non-negative.

Generally speaking, the energy consumption increases when the ambient temperature deviates in either direction from some desired value, say $\theta^{\text{des}}$ that is considered to be most comfortable to the occupants of a residential unit. Under these circumstances a temperature dependent term $d|\theta_k - \theta^{\text{des}}|$ should have been removed from $x_k^s$ so that the perceived utility is lowered in either direction by the difference between the temperature from the desired value. However, as the specific data used in this research involved only warmer days, it is assumed that $\theta_k > \theta^{\text{des}}$ thereby justifying the inclusion of the linear term $d_k \theta_k$ in (2). This approach also avoids using the non-differentiable absolute operator $|\cdot|$.

The derivative of $u(\mathbf{x}^s)$ with respect to any $x_k^s$ is the agent's marginal utility, which is also equal to the unit cost $c_k$. Accordingly, it can be readily established using (2) that,

$$\hat{x}_k^s = c_k^{-1} a_k + b_k + d_k \theta_k. \quad (3)$$

The above expression splits the model shiftable load $\hat{x}_k^s$ into cost dependent, constant, and temperature dependent terms. Fig. 2. illustrates how the utility and its derivative, the unit cost, vary with energy use. The utility is strictly concave in accordance with the law of diminishing returns. Although the unit cost is shown as a function in Fig. 2., in energy auctions the unit cost is usually determined by the DSO or aggregator, with each consumer agent responding by using the corresponding amount of energy [26],[27],[28],[29].

### B. Modeling Steps

The load matrix $\mathbf{X}$, the temperature matrix $\mathbf{\Theta}$, and the vector $\mathbf{c}$ of unit costs are the three input data sets. Load disaggregation is carried out using the hybrid approach proposed here, with the GMM associated with shiftable loads and NMF with fixed loads. The hybrid NMF-GMM disaggregation algorithm produces matrices of fixed loads $\tilde{\mathbf{X}}^f$, and shiftable loads $\tilde{\mathbf{X}}^s$, which serve as the input to the regression algorithm that yields the parameters associated with the consumer utility model. Fig. 3. illustrates the various steps involved.

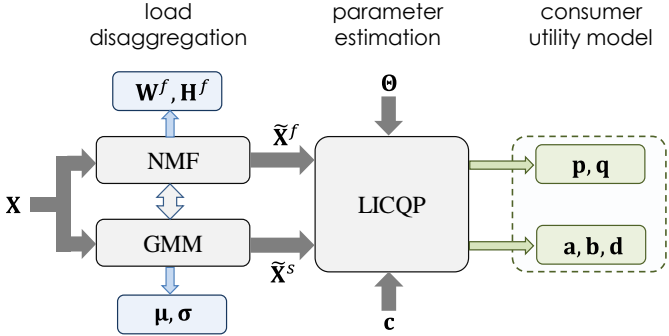

**Fig. 3.** Steps involved in the proposed approach. Thicker grey arrows indicate the directions of data flow whereas thinner, colored arrows represent parameter flow.

## III. Load Disaggregation

### A. Gaussian Mixture Model

The shiftable loads are represented in our framework in terms of a Gaussian mixture model (GMM). The $j^{\text{th}}$ prior, mean, and variance are $\pi_j$, $\mu_j$, and $\sigma_j^2$, which are trainable parameters with the exception of $\mu_0$ which is permanently assigned a value $\mu_0 = 0$ to represent the case when none of the shiftable loads is in use. Its variance $\sigma_0^2$ is treated as a nonzero quantity to subsume noise or small fixed loads. With $\mathcal{G}$ being the set of nonzero Gaussians, the probability distribution of $x_t^s$ is expressed as,

$$\text{pr}[x_t^s] = \sum_{j \in \mathcal{G} \cup \{0\}} \pi_j e^{-\frac{(x_t^s - \mu_j)^2}{\sigma_j^2}}. \quad (4)$$

The GMM is trained using the expectation-maximization (EM) algorithm [1], [33], briefly described here for convenience. The algorithm consists of an expectation step (E-step) and a maximization step (M-step). In the E-step, using existing estimates of $\pi_j$, $\mu_j$, and $\sigma_j^2$ for each $j$, the posterior probabilities $z_{j,t}(n) = \text{pr}[\delta_{g(t)=j}|x_t]$ of each Gaussian in $\mathcal{G} \cup \{0\}$ is estimated as,

$$z_{j,t}(n) = \Sigma_{t,n}^{-1} \pi_j e^{-\frac{(x_t(n) - \mu_j)^2}{\sigma_j^2}}, \quad j \in \mathcal{G} \cup \{0\}. \quad (5)$$

The quantity $\Sigma_{t,n}$ in the denominator of (5) is for normalization such that the joint probability is unity, $\sum_{j=0}^{|\mathcal{G}|} z_{j,t}(n) = 1$.

During the maximization step the GMM parameters, $\pi_j$, $\mu_j$, and $\sigma_j^2$ are updated using the following rules,

$$\mu_j = \sum_{t \in \mathcal{T}, n \in \mathcal{N}} z_{j,t} x_t(n), \quad j \in \mathcal{G}, \quad (6a)$$

$$\sigma_j^2 = \sum_{t \in \mathcal{T}, n \in \mathcal{N}} z_{j,t}(x_t(n) - \mu_j)^2, \quad j \in \mathcal{G} \cup \{0\}, \quad (6b)$$

$$\pi_j = |\mathcal{T}|^{-1}|\mathcal{N}|^{-1} \sum_{t \in \mathcal{T}, n \in \mathcal{N}} z_{j,t}(n), \quad j \in \mathcal{G} \cup \{0\}. \quad (6c)$$

The E-step and M-step are repeated multiple times until convergence.

Given a total load $x_t(n)$ at time instance $t$, the shiftable load is estimated according to the expression below,

$$\tilde{x}_t^s(n) = \max_{j \in \mathcal{G}} \mu_j \leq x_t(n). \quad (7)$$

Ignoring noise, the remaining load $x_t(n) - x_t^s(n)$ is taken to be the fixed load $\tilde{x}_t^f(n)$, i.e.,

$$\tilde{\mathbf{X}}^f = \mathbf{X} - \tilde{\mathbf{X}}^s. \quad (8)$$

The estimated fixed load $\tilde{\mathbf{X}}^f$ is disaggregated in the following NMF algorithm.

### B. Non-Negative Matrix Factorization

It is assumed that the fixed load $\tilde{x}^f(n)$ of each sample $n \in \mathcal{N}$ can be represented using a set $\mathcal{B}$ of basis vectors of dimensionality $|\mathcal{T}| \times 1$. If $h_k(n), k = 1, \ldots, |\mathcal{B}|$ are there coefficients, we must have,

$$\tilde{\mathbf{x}}^f(n) = \sum_{k \in \mathcal{B}} h_k^f(n) \mathbf{w}_k^f. \quad (9)$$

The basis set is assumed to be orthonormal, so that $\mathbf{w}_k^T \mathbf{w}_l = \delta_{k=l}$. These basis vectors form the columns of a $|\mathcal{T}| \times |\mathcal{B}|$ basis matrix, $\mathbf{W}^f = [h_k(n)]_{k \in \mathcal{K}}$. Similarly, letting $\mathbf{h}^f(n) = [h_k^f(n)]_{k \in \mathcal{K}}$ be the $B \times 1$ vector of coefficients in (9), we define the $|\mathcal{B}| \times |\mathcal{N}|$ matrix, $\mathbf{H}^f = [\mathbf{h}^f(n)]_{n \in \mathcal{N}}$. Therefore the above expression can be written more concisely as,

$$\tilde{\mathbf{X}}^f = \mathbf{W}^f \mathbf{H}^f. \quad (10)$$

As the basis vectors are mutually orthonormal, it is apparent that $\mathbf{W}^{f^T} \mathbf{W}^f = \mathbf{I}_{B \times B}$. The set of all possible orthonormal matrices $\mathbf{W}^f$ is a Steifel manifold, a subset of $\mathbb{R}^{|\mathcal{T}| \times |\mathcal{B}|}$, as long as $|\mathcal{B}| < |\mathcal{T}|$. Steifel manifolds are Riemannian under an appropriately defined matrix inner product [1], [34].

The matrices $\mathbf{W}^f$ and $\mathbf{H}^f$ are trained using the usual multiplicative update rule [17], [35], [36], [37] to minimize the squared Frobenius norm $\Phi = \|\mathbf{X}^f - \tilde{\mathbf{X}}^f\|_F^2$. Consider any matrix parameter $\mathbf{P}$ (which can be either $\mathbf{W}^f$ or $\mathbf{H}^f$). The gradient, $\nabla_\mathbf{P} \Phi$ of $\Phi$ can be expressed in terms of its positive and negative components as $\nabla_\mathbf{P} \Phi = \nabla_\mathbf{P}^+ - \nabla_\mathbf{P}^-$. In gradient descent,

**P** would be updated additively as $\mathbf{P} \leftarrow \mathbf{P} - \epsilon\nabla_\mathbf{P}^+ + \epsilon\nabla_\mathbf{P}^-$. This would require the use of an additional algorithmic constant $\epsilon$, as well as an additional projection step to ensure that $\mathbf{P} \geq \mathbf{0}$. The multiplicative method removes both these drawbacks. The multiplicative update rule is $\mathbf{P} \leftarrow \mathbf{P} \circ \nabla_\mathbf{P}^- \oslash \nabla_\mathbf{P}^+$.

Since the gradient $\nabla_{\mathbf{H}^f}\Phi = \mathbf{W}^{f^T}\mathbf{W}^f\mathbf{H}^f - \mathbf{W}^{f^T}\mathbf{X}^f$, the coefficient matrix $\mathbf{H}^f$ is updated according to the following expression,

$$\mathbf{H}^f \leftarrow \mathbf{H}^f \circ \frac{\mathbf{W}^{f^T}\mathbf{X}^f}{\mathbf{W}^{f^T}\mathbf{W}^f\mathbf{H}^f}. \quad (11)$$

The corresponding update rule for the basis matrix $\mathbf{W}^f$ could be obtained in the same manner. However, in order to maintain orthonormality of the basis vectors, the version in [38] has been adopted in this research, which yields the modified update rule,

$$\mathbf{W}^f \leftarrow \mathbf{W}^f \circ \frac{\mathbf{X}^f \mathbf{H}^{f^T}}{\mathbf{W}^f \mathbf{H}^f \mathbf{X}^{f^T} \mathbf{W}^f}. \quad (12)$$

The modification lies in the denominator of the above rule. Since the gradient $\nabla_{\mathbf{W}^f}\Phi$ in $\mathbb{R}^{|\mathcal{T}|\times|\mathcal{B}|}$ is $\mathbf{W}^f\mathbf{H}^f\mathbf{W}^{f^T} - \mathbf{X}^f\mathbf{H}^{f^T}$, the denominator in (12) would have been $\mathbf{W}^f\mathbf{H}^f\mathbf{W}^{f^T}$. However its natural gradient $\widetilde{\nabla}_{\mathbf{W}^f}\Phi$ must be tangential to the "curved" Steifel manifold in $\mathbb{R}^{|\mathcal{T}|\times|\mathcal{B}|}$, which is possible as long as $\mathbf{W}^{f^T}\widetilde{\nabla}_{\mathbf{W}^f} + \widetilde{\nabla}_{\mathbf{W}^f}^T\mathbf{W}^{f^T} = \mathbf{0}$ [34]. Accordingly, the natural gradient of $\Phi$ is $\widetilde{\nabla}_{\mathbf{W}^f}\Phi = \mathbf{W}^f\mathbf{H}^f\mathbf{X}^{f^T}\mathbf{W}^f - \mathbf{X}^f\mathbf{H}^{f^T}$, leading to the update rule in (11b) that preserves the orthonormality of the basis matrix $\mathbf{W}^f$.

It must be mentioned here that although it has been shown in [36] that the multiplicative update rule lowers the cost function $\Phi$, to the best of our knowledge there have been no concrete studies showing that it converges to a unique fixed point. Needless to say, the simplicity of multiplicative updates makes it widely used. The theoretical backdrop becomes even more pessimistic with orthonormal NMF updates as applied in this research. Nonetheless, numerous preliminary simulations in this research revealed that this method was consistently as accurate as that in [35], while maintaining orthonormality.

Ignoring noise, the residual load in $\mathbf{X}$ that is not fixed, is treated as the shiftable load, so that,

$$\widetilde{\mathbf{X}}^s = \mathbf{X} - \mathbf{W}^f\mathbf{H}^f. \quad (13)$$

This load is used by the next iteration of the EM algorithm.

### C. Hybrid Algorithm

The steps of the hybrid algorithm are outlined below.

1. **initialize** $\widetilde{\mathbf{X}}^f, \widetilde{\mathbf{X}}^s$
2. **initialize** $\{\pi_j, \mu_j, \sigma_j^2\}_{j \in \mathcal{G} \cup \{0\}}$
3. **initialize** $\mathbf{W}^f$, $\mathbf{H}^f$
4. **repeat**
   a. **estimate** $\{z_{j,t}(n)\}_{(j,t,n) \in \mathcal{G}\times\mathcal{T}\times\mathcal{N}}$ according to (4),(5)
   b. **update** $\{\pi_j, \mu_j, \sigma_j^2\}_{j \in \mathcal{G}\cup\{0\}}$ according to (6)
   c. **estimate** $\widetilde{\mathbf{X}}^f$ according to (7),(8)
   d. **update** $\mathbf{H}^f$ according to (11)
   e. **update** $\mathbf{W}^f$ according to (12)
   f. **estimate** $\widetilde{\mathbf{X}}^s$ according to (13)
   **until** converged

The first step of the hybrid disaggregation algorithm involves dividing the training samples $\mathbf{X}$ into shiftable and fixed components. Initialization of the GMM parameters and the NFM matrices takes place in steps 2 and 3. Step 4 is the main iterative process.

Steps 4a and 4b pertain to the GMM as outlined in III.A. In step 4a, the hidden variables $z_{j,t}(n)$ are estimated. This is the E-step of the EM algorithm. The M-step is implemented in step 4b. In step 4c, the estimated fixed load $\widetilde{\mathbf{X}}^f$ is computed. The next two steps (steps 4d and 4e) implement the NMF algorithm. In step 4d and 4e, the matrices $\mathbf{H}^f$ and $\mathbf{W}^f$ associated with the fixed loads are incremented. Finally, $\widetilde{\mathbf{X}}^s$, the estimate of the total shiftable load is determined in step 4f.

The criterion to determine when convergence is achieved is omitted in the overall algorithm's outline because it was found that the algorithm always converged within merely 10-20 iterations, which was well within 2 minutes. Therefore the repeat-until loop may be replaced with a simple for loop that terminates after a predetermined maximum number of iterations.

It may be noted that in (8), the shiftable load $\widetilde{\mathbf{X}}^s$, which is determined using a probability threshold, is subtracted from the total load $\mathbf{X}$ for the NMF steps. In a similar manner, the NMF algorithm estimates $\widetilde{\mathbf{X}}^f$ and the difference $\mathbf{X} - \widetilde{\mathbf{X}}^f$ is used by the EM algorithm. Suppose the estimated total load, $\widetilde{\mathbf{X}}$ is given by,

$$\widetilde{\mathbf{X}} = \widetilde{\mathbf{X}}^s + \widetilde{\mathbf{X}}^f. \quad (13)$$

The difference between $\mathbf{X}$ and $\widetilde{\mathbf{X}}$ can be attributed to either noise or discrepancy arising from the proposed hybrid disaggregation algorithm. The disaggregated loads $\widetilde{\mathbf{X}}^s$ and $\widetilde{\mathbf{X}}^f$ are used to estimate the model parameters, which is described next.

## IV. PARAMETER ESTIMATION

### A. $\epsilon$-Insensitive Constraints

The $|\mathcal{K}| \times 1$ vectors of parameters are, $\mathbf{a}, \mathbf{b}, \mathbf{d}, \mathbf{p}$, and $\mathbf{q}$. Let $\epsilon^f$ be a tolerance bound on the difference between the modeled and the disaggregated fixed loads in (1) and (9). No penalty is incurred when the absolute difference, $|\hat{x}_k^f - \tilde{x}_k^f|$ is within $\epsilon^f$, whereas anything in excess of it is an error, $\xi_k^f$. Hence, for each sample $n \in \mathcal{N}$ is,

$$\mathbb{I}^f: |\mathbf{p} + \boldsymbol{\theta}(n) \circ \mathbf{q} - \mathbf{G}\widetilde{\mathbf{x}}^f(n)| \leq \mathbf{1}_{|\mathcal{K}|}\epsilon^f + \boldsymbol{\xi}^f. \quad (14)$$

In a similar manner, with $\epsilon^s$ being the maximum allowable tolerance of the difference between the model and the disaggregated shiftable loads, shown in (3) and (10) for each $n \in \mathcal{N}$ is given by,

$$\mathbb{I}^s: |\mathbf{c}^{\circ-1} \circ \mathbf{a} + \mathbf{b} + \boldsymbol{\theta}(n) \circ \mathbf{d} - \mathbf{G}\widetilde{\mathbf{x}}^s(n)| \leq \mathbf{1}_{|\mathcal{K}|}\epsilon^s + \boldsymbol{\xi}^s. \quad (15)$$

In (14) and (15), $\mathbf{G}$ is a $|\mathcal{K}| \times |\mathcal{T}|$ summing matrix that sums time samples of variables index $t$ to their periodic totals indexed $k$, as shown,

$$\mathbf{G} = \begin{bmatrix} \mathbf{1}_L^T & \cdots & \mathbf{0}_L^T \\ \vdots & \ddots & \vdots \\ \mathbf{0}_L^T & \cdots & \mathbf{1}_L^T \end{bmatrix}.$$

The use of $\epsilon$-insensitive loss as in (14) and (15) is universally accepted practice in machine learning as it reduces the regression model's VC-dimensionality [39], [40].

*B. $L_2$ Regularized Loss*

The weighted sum of all squared errors $\xi_k^f$ and $\xi_k^s$, is the loss function is selected as the loss function to be minimized. This loss, $\mathbb{L}$ can be expressed as a squared $L_2$ norm,

$$\mathbb{L} = \left\| \begin{bmatrix} w^f \mathbf{I} & \mathbf{0} \\ \mathbf{0} & w^s \mathbf{I} \end{bmatrix} \begin{bmatrix} \boldsymbol{\xi}^f \\ \boldsymbol{\xi}^s \end{bmatrix} \right\|^2. \quad (16)$$

The quantities $w^f$ and $w^s$ are the weights.

The above loss is regularized by with the function which is the sum of two squared $L_2$ norms of the shiftable and fixed load parameters. The regularization function $\mathbb{R}$ is given by the following expression,

$$\mathbb{R} = \left\| \begin{bmatrix} \gamma^p \mathbf{I} & \mathbf{0} \\ \mathbf{0} & \gamma^q \mathbf{I} \end{bmatrix} \begin{bmatrix} \mathbf{p} \\ \mathbf{q} \end{bmatrix} \right\|^2 + \left\| \begin{bmatrix} \gamma^a \mathbf{I} & \mathbf{0} & \mathbf{0} \\ \mathbf{0} & \gamma^b \mathbf{I} & \mathbf{0} \\ \mathbf{0} & \mathbf{0} & \gamma^d \mathbf{I} \end{bmatrix} \begin{bmatrix} \mathbf{a} \\ \mathbf{b} \\ \mathbf{d} \end{bmatrix} \right\|^2. \quad (17)$$

In (17), $\gamma^p$, $\gamma^q$, $\gamma^a$, $\gamma^b$ and $\gamma^d$ are arbitrarily small, positive regularization coefficients.

*C. Linear Inequality Constrained Quadratic Programming*

The overall problem of estimating the parameter vectors, $\mathbf{a}, \mathbf{b}, \mathbf{d}, \mathbf{p}$, and $\mathbf{q}$ of the utility model can now be readily formulated in terms of the following linear inequality constrained quadratic programming (LICQP),

Minimize: $\mathbb{L} + \mathbb{R}$.
Subject to: $\mathbb{I}^f, \mathbb{I}^s$.

It can be observed that the regularized loss can be expressed as a quadratic function with an associated $7|\mathcal{K}| \times 7|\mathcal{K}|$ matrix whose diagonal elements are the two weights and the five coefficients; ergo the quadratic term in the objective function $\mathbb{L} + \mathbb{R}$ is strictly positive definite. Thus, the objective function in the above LICQP is strictly convex. As both set of inequality constraints $\mathbb{I}^f$, $\mathbb{I}^s$ are linear, there exists a unique minimum.

## V. RESULTS & DISCUSSION

*A. Data*

The proposed approach was tested on actual energy usage data of two residential customers that was obtained from the Pecan Street Inc., Dataport database [41] sampled at one minute intervals, and for 61 days during March and April, 2018. The total energy data was arranged as a $1440 \times 61$ data matrix $\mathbf{X}$ whose columns were the 1440-dimensional sample vectors, $\mathbf{x}(n), n \in \mathcal{N}$.

The database also included individual appliance usage measurements at the circuit level. These appliances are, (*i*) air conditioner, (*ii*) electric car, (*iii*) washer/dryer, (*iv*) dishwasher, (*v*) microwave, (*vi*) refrigerator, (*vii*) furnace, (*viii*) bedroom appliances, (*ix*) light plugs and (*x*) kitchen appliances. The first four appliances are classified as shiftable loads for our purpose, while the remaining ones are fixed loads. It should be noted that user-1 did not own bedroom appliances or furnace, while user-2 did not own an electric car.

The hourly temperature data for the same 61 days $n \in \mathcal{N}$ was also obtained from [41]. Therefore, the load data and temperatures in the data used in this study are from the city of Austin in Texas. Accordingly, the hourly time of use rates were obtained from the Austin Energy website [42]. This site provides electricity rates in the Austin area for a time of use pilot project. The user data used in this study were part of DER that received pricing information from the utility and were expected to defer their loads during peak or high energy price hours. All algorithms were implemented in MATLAB.

*B. Disaggregation*

The plots in Figs. 4. and 5. show comparisons between the real data (left column, red) and the outcome of disaggregation for user-1 (right column, blue) for the first two days, ($n = 1, 2$). The remarkable similarities in the shiftable loads $\mathbf{X}^s$ and $\widetilde{\mathbf{X}}^s$ (top row) is evident. The corresponding fixed loads $\mathbf{X}^f$ and $\widetilde{\mathbf{X}}^f$ (middle row) are quite similar but there are minor differences.

Among the most conspicuous ones in Fig. 4 occur at time instances $t \approx 250, 660, 718$, and $t \approx 1080$. At $t \approx 250, 660$, a small fraction of the fixed load $\mathbf{X}^f$ is erroneously incorporated by the hybrid algorithm into the shiftable load $\widetilde{\mathbf{X}}^s$, whereas at $t \approx 718, 1080$ some of $\mathbf{X}^s$ is transferred to $\widetilde{\mathbf{X}}^f$. However these discrepancies are quite small in spite of their exaggerated appearances in the plots are due to the smaller range of the fixed loads in comparison to the shiftable loads.

In Fig. 5, between , $t \approx 660 - 800$ , some differences between the real and proposed approach's reconstructed shift and shiftable loads can be observed. In particular, the high frequency fluctuation in $\mathbf{X}^f$ does not get incorporated in $\widetilde{\mathbf{X}}^f$. However, this is at only at the relatively low power level of 0.5 kW. In the interval $t \approx 820 - 900$, a few upward spikes in $\widetilde{\mathbf{X}}^f$ can be seen. This occurs again at the instances $t \approx 1210, 1380$.

In both Fig. 4 and Fig. 5, the total loads $\mathbf{X}$ and $\widetilde{\mathbf{X}}$ (bottom rows) are very similar, indicating the effectiveness of the hybrid approach with the NMF and GMM algorithms working in tandem in disaggregating the total load.

Figs. 6 and 7 show the consumption patterns of user-2 on days $n = 1, 2$. The positions of the real loads (left, red), disaggregated loads (right, blue), shiftable loads (top), fixed loads (middle) and total (bottom) are exactly as in Figs. 4 and 5 which were associated with user-1.

One noticeable difference in user-2 consumption profile is that the power usages of the fixed loads and shiftable loads are at similar levels. Consequently, sudden bursts in the fixed loads were consistently misinterpreted by the hybrid disaggregation algorithm as shiftable loads, with the GMM assigning a prior mean $\mu_j$ (for some $j$), with a variance $\sigma_j^2$ large enough to subsume all such fixed load spikes. These spikes appear at time instances $t \approx 491, 600, 776$ on day $n = 1$ and at $t \approx 770, 481, 485, 1210, 1360$ on day $n = 2$. Although this phenomenon has been observed earlier in the case of user-1, it is visually more pronounced now due to the similar power levels of user-2's shiftable and fixed loads.

Simple post-processing heuristic filtering would be readily able to rectify these discrepancies in the hybrid disaggregation approach. In view of algorithmic simplicity, no modifications were made to the disaggregation approach to address this situation. Moreover, as these fixed load spikes lasted for only 1-2 minutes each, the effect on model parameter estimation is imperceptible.

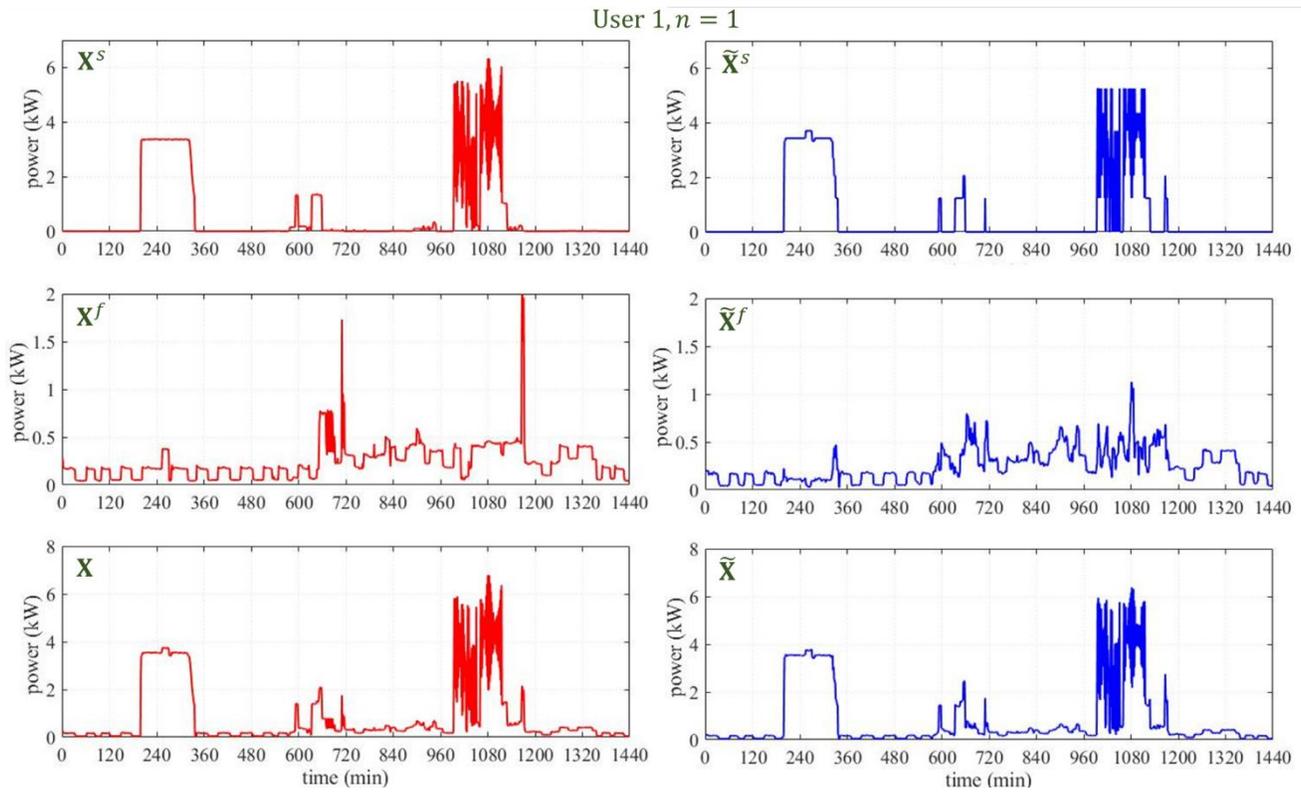

**Fig. 4.** Energy consumption of user-1 on day $n = 1$. The real loads (red, left) and disaggregated loads (blue, right) are shown. The shiftable (top) and fixed (middle) components of the total load (bottom) are shown.

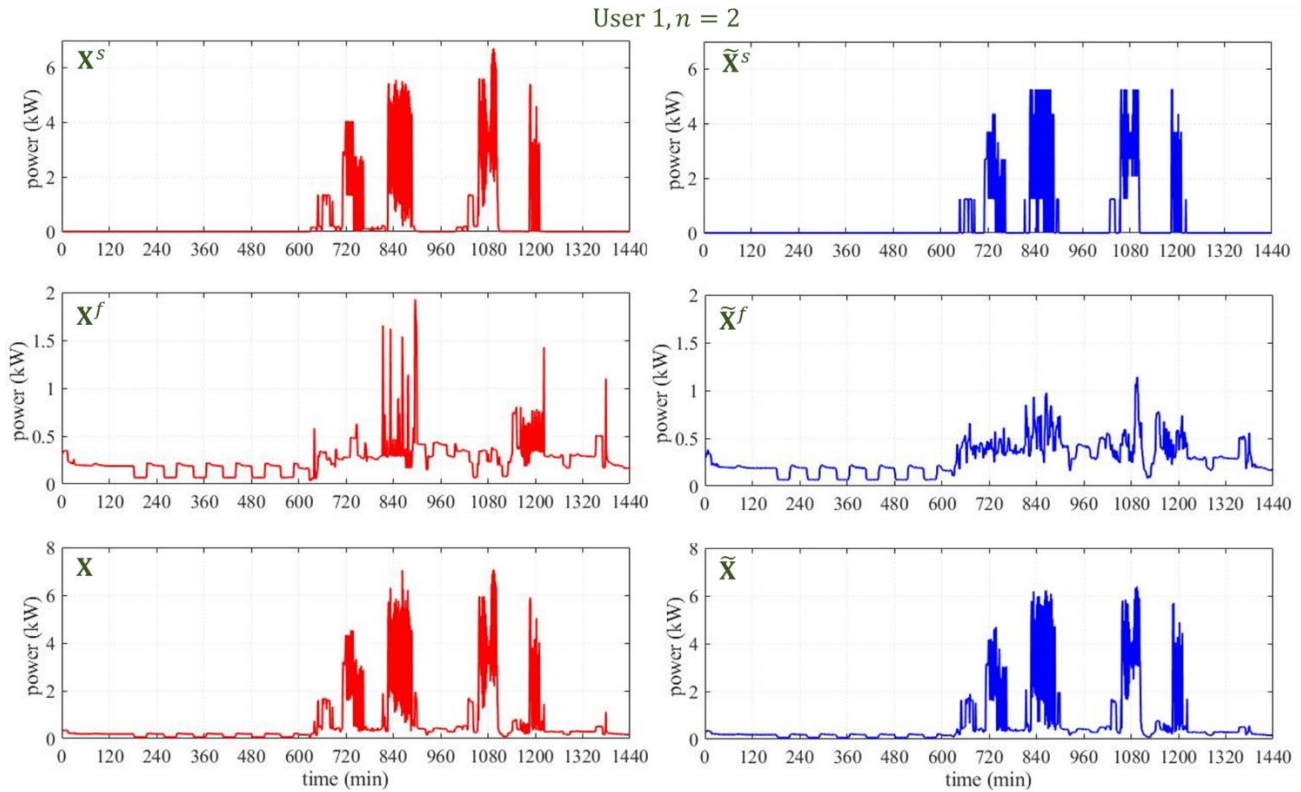

**Fig. 5.** Energy consumption of user-1 on day $n = 2$. The real loads (red, left) and disaggregated loads (blue, right) are shown. The shiftable (top) and fixed (middle) components of the total load (bottom) are shown.

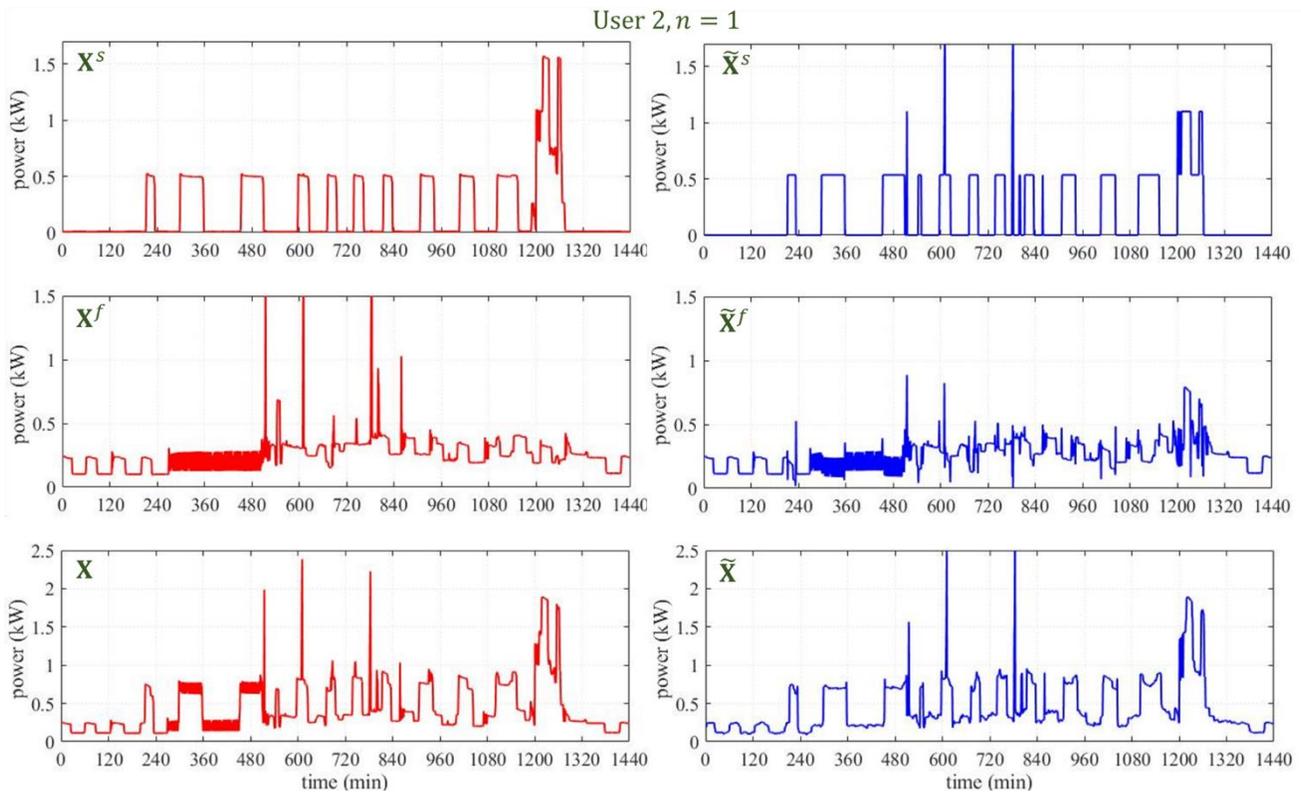

**Fig. 6.** Energy consumption of user-1 on day $n = 1$. The real loads (red, left) and disaggregated loads (blue, right) are shown. The shiftable (top) and fixed (middle) components of the total load (bottom) are shown.

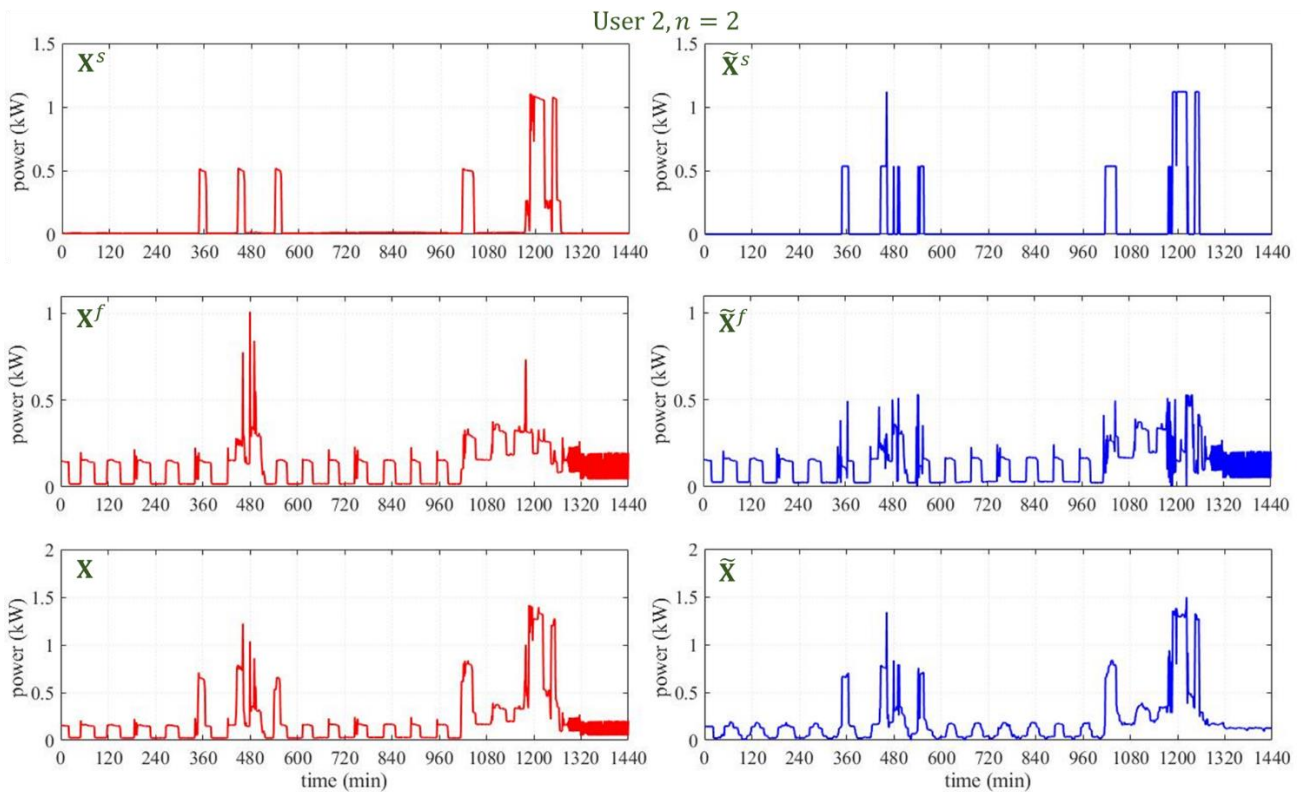

**Fig. 7.** Energy consumption of user-1 on day $n = 1$. The real loads (red, left) and disaggregated loads (blue, right) are shown. The shiftable (top) and fixed (middle) components of the total load (bottom) are shown.

## C. Parameter Estimation

The parameter vectors, $\mathbf{a}, \mathbf{b}, \mathbf{d}, \mathbf{p}$, and $\mathbf{q}$ of the utility model are the optimal solutions of the LICQP in IV.C, that were obtained using the fmincon() subroutine that is available in MATLAB's optimization toolbox. As it was noticed that temperature had little bearing on the fixed load usage of either user, the corresponding parameter $\mathbf{q}$ was set to zero. Although the purpose of this research is to derive a consumer model entirely through smart metering data, sans any prior information about how and when individual appliances were operated, $\mathbf{a}, \mathbf{b}, \mathbf{d}, \mathbf{p}$, and $\mathbf{q}$ were computed using real loads $\mathbf{X}^s$ and $\mathbf{X}^f$ as well as their estimates $\widetilde{\mathbf{X}}^s$ and $\widetilde{\mathbf{X}}^f$ obtained as described in V.B using the hybrid approach. The median values of all parameters are supplied in Table I, separately for each user. It can be seen that the medians obtained from the disaggregated estimates (4th column) are very close to those obtained from real data (3rd column). Furthermore, the correlation coefficients between the median values are also shown in Table I (5th column). The correlation coefficients of the shiftable load parameters $\mathbf{a}, \mathbf{b}$ and $\mathbf{d}$ are in the range 0.978 – 0.995, which is remarkably close to 1, indicating the fidelity of the proposed disaggregation approach. The corresponding values of the fixed load parameter $\mathbf{p}$ (0.891 and 0.892) are also very close to 1.

TABLE I.

| user | para-meter | median | | corr. coeff. |
|---|---|---|---|---|
| | | real | disagg. | |
| User-1 | a | 2.816 | 2.711 | 0.994 |
| | b | 0.222 | 0.242 | 0.983 |
| | d | 0.219 | 0.236 | 0.983 |
| | p | 27.191 | 20.565 | 0.891 |
| | q | 0 | 0 | - |
| User-2 | a | 2.819 | 2.709 | 0.995 |
| | b | 0.228 | 0.269 | 0.979 |
| | d | 0.225 | 0.263 | 0.978 |
| | p | 27.192 | 21.108 | 0.892 |
| | q | 0 | 0 | - |

The utility curves associated with the shiftable loads were obtained from the parameters $\mathbf{a}, \mathbf{b}$ and $\mathbf{d}$ are shown in Fig. 8 for user-1 and Fig. 9 for user-2. In both figures, the users' perceived utilities (top) as well as their derivatives (bottom) are provided. These plots were obtained using parameters obtained directly from real loads $\mathbf{X}^s$ and $\mathbf{X}^f$ (red, solid lines) as well as their estimates $\widetilde{\mathbf{X}}^s$ and $\widetilde{\mathbf{X}}^f$ (blue, dotted lines). It should be noted that the difference between the real and utilities is only an artifact of integration. The curves corresponding to the estimated utilities (blue, dotted) can be shifted upwards to more closely with the real curves (red, solid).

Finally, Fig. 10 shows the hourly consumption profiles of user-1 (top) and user-2 (bottom) for a typical day in the 61 day period used in this research. They were obtained from (1) and (2) with the parameters obtained from the LICQP optimization procedure and using the disaggregated loads estimated by the disaggregation approach.

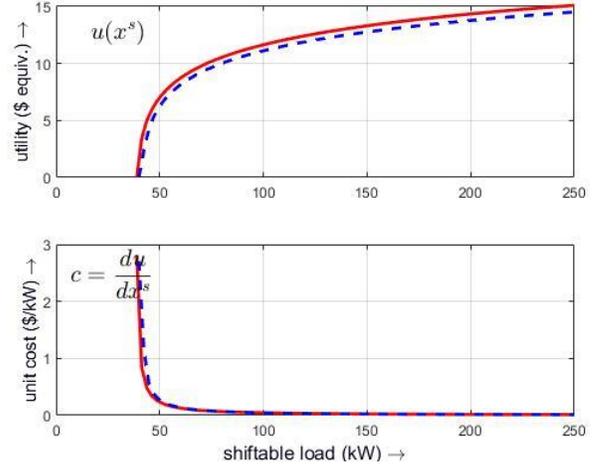

**Fig. 8.** User-1 utility and cost. The utility functions (top) and cost (bottom) functions are shown, using real loads (red, solid lines) and disaggregated loads (blue dotted lines).

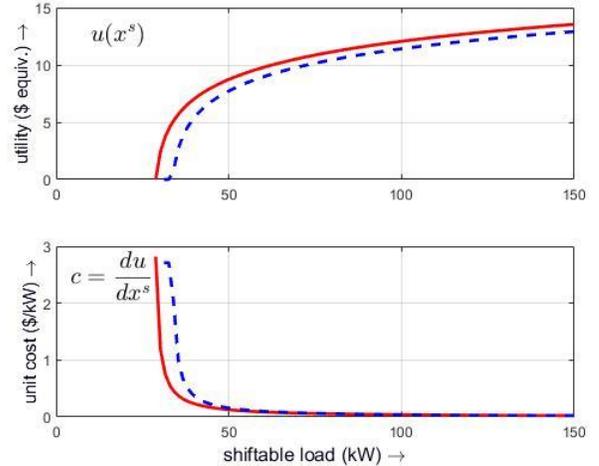

**Fig. 9.** User-2 utility and cost. The utility functions (top) and cost (bottom) functions are shown, using real loads (red, solid lines) and disaggregated loads (blue dotted lines).

## VI. CONCLUSION

This paper proposes a generic class of data-driven semiparametric models derived from consumption data of residential consumers. A two-stage machine learning approach is developed, in which disaggregation of the load into fixed and shiftable components is accomplished by means of a hybrid algorithm consisting of non-negative NMF and GMM, with the latter trained by an expectation-maximization (EM) algorithm. In the second stage, the model parameters are estimated using an $L_2$-norm, $\epsilon$-insensitive regression approach. Actual energy usage data of two residential customers show the validity of the proposed method.

The proposed method that disaggregates electricity consumption into the shiftable and fixed portions, can enhance the demand response program of utilities as well as help residential customers make well informed decisions and increase the energy efficiency gains while participating in a demand response program. It also allows for automated energy transactions between customers and the utilities based on a semi-parametric utility model learned from the actual consumption pattern of the customers.

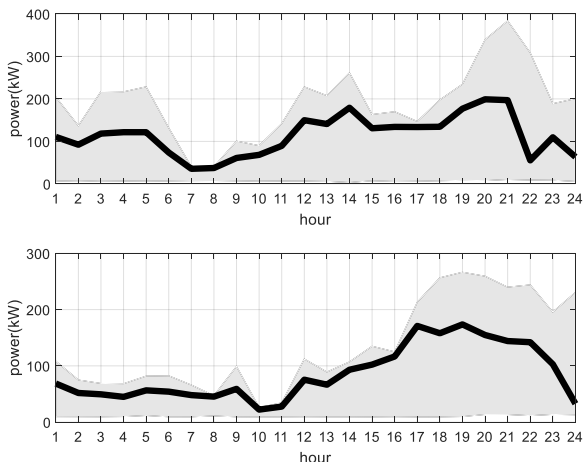

**Fig. 10.** Typical 24-hour load profiles of user-1 (top) and user-2 (bottom) obtained from the semi-parametric utility model with disaggregated load estimates. The shaded regions depict the range of values from the real data.

## VII. Acknowledgment

This work was supported by the National Science Foundation-CPS under Grant CNS-1544705.